\documentclass[aps,prl,twocolumn,showpacs,superscriptaddress]{revtex4}  
\usepackage{graphicx}  
\usepackage{dcolumn}   
\usepackage{bm}        
\usepackage{amssymb}   

\input babarsym

\newcommand{\BABARPubYear}    {07}
\newcommand{\BABARPubNumber}  {071}
\newcommand{\SLACPubNumber} {13056}

\begin{document}

\begin{flushleft}
\babar-PUB-\BABARPubYear/\BABARPubNumber \\
SLAC-PUB-\SLACPubNumber \\
\end{flushleft}

\title{A Measurement of the Branching Fractions of exclusive {\boldmath $\Bbar \to D^{(*)}(\pi) \ell^- \bar{\nu}_{\ell}$} Decays in Events with a Fully Reconstructed {\boldmath $B$} Meson}

%
\author{B.~Aubert}
\author{M.~Bona}
\author{Y.~Karyotakis}
\author{J.~P.~Lees}
\author{V.~Poireau}
\author{X.~Prudent}
\author{V.~Tisserand}
\author{A.~Zghiche}
\affiliation{Laboratoire de Physique des Particules, IN2P3/CNRS et Universit\'e de Savoie, F-74941 Annecy-Le-Vieux, France }
\author{J.~Garra~Tico}
\author{E.~Grauges}
\affiliation{Universitat de Barcelona, Facultat de Fisica, Departament ECM, E-08028 Barcelona, Spain }
% ...
\collaboration{The \babar\ Collaboration}
\noaffiliation

\date{\today}

\begin{abstract}
We report a measurement of the branching fractions for $\Bbar \to D^{(*)}(\pi) \ell^- \bar{\nu}_{\ell}$ decays based on 
341.1 fb$^{-1}$ of data collected at the $\Upsilon(4S)$ resonance with the \babar\
detector at the \pep2\ $e^+e^-$ storage rings. 
Events are tagged by fully reconstructing one of the $B$ mesons in a hadronic decay mode. We obtain ${\cal B} (B^- \to D^0 \ell^- \bar{\nu}_{\ell}) = (2.33 \pm 0.09_{stat.} \pm 0.09_{syst.})\%$, ${\cal B} (B^- \to D^{*0} \ell^- \bar{\nu}_{\ell}) = (5.83 \pm 0.15_{stat.} \pm 0.30_{syst.})\%$, ${\cal B} (\Bzb \to D^+ \ell^- \bar{\nu}_{\ell}) = (2.21 \pm 0.11_{stat.} \pm 0.12_{syst.})\%$, ${\cal B} (\Bzb \to D^{*+} \ell^- \bar{\nu}_{\ell}) = (5.49 \pm 0.16_{stat.} \pm 0.25_{syst.})\%$, ${\cal B} (B^- \to D^+ \pi^- \ell^- \bar{\nu}_{\ell}) = (0.42 \pm 0.06_{stat.} \pm 0.03_{syst.})\%$, ${\cal B} (B^- \to D^{*+}\pi^- \ell^- \bar{\nu}_{\ell}) = (0.59 \pm 0.05_{stat.} \pm 0.04_{syst.})\%$, ${\cal B} (\Bzb \to D^0 \pi^+ \ell^- \bar{\nu}_{\ell}) = (0.43 \pm 0.08_{stat.} \pm 0.03_{syst.})\%$ and ${\cal B} (\Bzb \to D^{*0} \pi^+ \ell^- \bar{\nu}_{\ell}) = (0.48 \pm 0.08_{stat.} \pm 0.04_{syst.})\%$.\end{abstract}

\maketitle 

The determination of the individual exclusive branching fractions of
$\Bbar \to X_c \ell^- \bar{\nu}_{\ell}$ decays~\cite{ell} is important for the study of the
 semileptonic decays of the $B$ meson. 
Improvement in the knowledge  of these branching fractions is also important to reduce the systematic uncertainty in the measurements of the Cabibbo-Kobayashi-Maskawa~\cite{CKM} matrix elements $|V_{cb}|$ and $|V_{ub}|$. For example, one of the leading sources of
systematic uncertainty in the extraction of $|V_{cb}|$ from the exclusive
decay $\Bbar \to D^* \ell^- \bar{\nu}_{\ell}$ is the limited knowledge of the
background due to $\Bbar \to D^* \pi \ell^- \bar{\nu}_{\ell}$. Improved measurements of $\Bbar \to X_c \ell^- \bar{\nu}_{\ell}$ decays will also benefit the accuracy of the extraction of $|V_{ub}|$, as analyses are
extending into kinematic regions in which these decays represent a sizable background.

Based on current measurements~\cite{aleph,delphi,babar-1,babar-2,belle} the rate of inclusive semileptonic $B$ decays exceeds the sum of the measured exclusive decay rates~\cite{pdg}. While $\Bbar \rightarrow D \ell^- \bar{\nu}_{\ell}$ and  $D^* \ell^- \bar{\nu}_{\ell}$ decays account for about 70\% of this total, the contribution of other states, including resonant and non-resonant $D^{(*)}\pi \ell^- \bar{\nu}_{\ell}$ (denoted by $D^{**} \ell^- \bar{\nu}_{\ell}$), is not yet well measured and may help to explain the inclusive-exclusive discrepancy. 
 
In this letter, we present measurements of the branching fractions for $\Bbar \to D^{(*)}(\pi) \ell^- \bar{\nu}_{\ell}$  decays, separately for charged and neutral $B$ mesons. 

The analysis is based on data collected with the \babar\ detector~\cite{detector} at the 
\pep2\ asymmetric-energy $e^+e^-$ storage rings. The data consist of a total 
of 341.1~fb$^{-1}$ recorded at the $\Upsilon(4S)$ 
resonance, corresponding to 378 million \BB\ pairs. An additional 36~fb$^{-1}$  off-peak data sample, taken at a center-of-mass (CM) energy 40 MeV below the $\Upsilon(4S)$ resonance, is used to study background from $e^+e^- \to f\bar{f}~(f=u,d,s,c,\tau)$ events (continuum production). A detailed GEANT4-based Monte Carlo (MC) simulation~\cite{Geant} of \BB\ and continuum events is used to study the detector response, its acceptance, and to test the analysis techniques. The simulation models $\Bbar \to D^{(*)} \ell^- \bar{\nu}_{\ell}$
decays using calculations based on Heavy Quark Effective Theory~\cite{HQETBaBar}, $\Bbar \to D^{**}(\rightarrow D^{(*)} \pi) \ell^- \bar{\nu}_{\ell}$ decays using the ISGW2 model~\cite{ISGW}, and $\Bbar
\to D^{(*)} \pi \ell^- \bar{\nu}_{\ell}$ decays using the Goity-Roberts model~\cite{Goity}.

We select semileptonic $B$ decays in events containing
a fully reconstructed $B$ meson ($B_{tag}$), which allows us to constrain the kinematics, reduce the combinatorial background, and determine the charge and flavor of the signal $B$ meson.

We first reconstruct the semileptonic $B$ decay, selecting a lepton with momentum $p^*_{\ell}$ in the  center-of-mass  frame higher than 0.6 GeV/$c$. Electrons from photon conversions and $\pi^0$ Dalitz decays are removed by searching for pairs of oppositely charged tracks that form a vertex with an invariant mass compatible with a photon conversion or a $\pi^0$ Dalitz decay.  Candidate $D^0$ mesons, having the correct 
charge-flavor correlation with the lepton, are reconstructed
in the $K^-\pi^+$, $K^- \pi^+ \pi^0$, $K^- \pi^+ \pi^+ \pi^-$,
$K^0_S \pi^+ \pi^-$, $K^0_S \pi^+ \pi^- \pi^0$, $K^0_S \pi^0$, $K^+ K^-$,
$\pi^+ \pi^-$, and $K^0_S K^0_S$ channels, and $D^+$ mesons in the
$K^- \pi^+ \pi^+$, $K^- \pi^+ \pi^+ \pi^0$, $K^0_S \pi^+$, $K^0_S \pi^+ \pi^0$,
$K^+ K^- \pi^+$, $K^0_S K^+$, and $K^0_S \pi^+ \pi^+ \pi^-$ channels.
In events with multiple $\Bbar \to D \ell^- \bar{\nu}_{\ell}$ candidates, the candidate with the best  $D$-$\ell$ vertex fit is selected.
Candidate $D^*$ mesons are reconstructed by combining a $D$ candidate with a pion or a photon in the $D^{*+} \rightarrow D^0 \pi^+ $, $D^{*+} \rightarrow D^+ \pi^0$, $D^{*0} \rightarrow D^0 \pi^0$, and $D^{*0} \rightarrow D^0 \gamma$ channels. 
In events with multiple $\Bbar \to D^{*} \ell^- \bar{\nu}_{\ell}$ candidates, we choose the candidate with the smallest $\chi^2$ based on the deviations from the nominal values of the $D$ invariant mass and the invariant mass difference between the $D^*$ and the $D$, using the measured resolution. 

 We reconstruct $B_{tag}$ decays of the type $\Bbar \rightarrow D Y$, where 
$Y$ represents a collection of hadrons with a total charge of $\pm 1$, composed
of $n_1\pi^{\pm}+n_2 K^{\pm}+n_3 K^0_S+n_4\pi^0$, where $n_1+n_2 \leq  5$, $n_3
\leq 2$, and $n_4 \leq 2$. Using $D^0(D^+)$ and $D^{*0}(D^{*+})$ as seeds for $B^-(\Bzb)$ decays, we reconstruct about 1000 different decay chains.

The kinematic consistency of a $B_{tag}$ candidate with a $B$ meson decay is evaluated using two variables: the beam-energy
substituted mass $m_{ES}=\sqrt{s/4-|p^*_B|^2}$, and the energy difference $\Delta E = E^*_B -\sqrt{s}/2$. Here $\sqrt{s}$ refers to the total CM  energy, and $p^*_B$ and $E^*_B$ denote the momentum and energy of the $B_{tag}$ candidate in the CM frame. For correctly identified $B_{tag}$ decays, the $m_{ES}$ distribution peaks at the $B$ meson mass, while $\Delta E$ is consistent
with zero.
We select a $B_{tag}$ candidate in the signal region
defined as 5.27~GeV/$c^2$ $< m_{ES} <$ 5.29~GeV/$c^2$, excluding $B_{tag}$ candidates with
 daughter particles in common with the charm meson or
the lepton from the semileptonic $B$ decay. In the case of multiple $B_{tag}$ candidates in an event, we select the one with the smallest
$|\Delta E|$ value. The $B_{tag}$ and the $D^{(*)}\ell$ candidates are required to have the correct charge-flavor correlation. Mixing effects in the $\Bzb$ sample are accounted for as described in~\cite{BBmixing}. 
Cross-feed effects, $i.e.$, $B^-_{tag} (\Bzb_{tag})$ candidates erroneously reconstructed as a neutral~(charged) $B$,  are subtracted using estimates from the simulation.

For $\Bbar \rightarrow D^{(*)} X \ell^- \bar{\nu}_{\ell}$ decays, $D(D^{*})$ candidates are selected within 2$\sigma$~(1.5-2.5$\sigma$, depending on the $D^*$ decay mode) of the $D$ mass ($D^{*}-D$ mass difference), with $\sigma$ typically around 8~(1-7) MeV$/c^{2}$. We also require the cosine of the angle between the directions of the $D^{(*)}$
candidate and the lepton in the CM frame to be less than zero, to reduce background from non-$B$ semileptonic decays.

We reconstruct $B^- \rightarrow D^{(*)+}\pi^- \ell^- \bar{\nu}_{\ell}$ and $\Bzb \rightarrow D^{(*)0}\pi^+ \ell^- \bar{\nu}_{\ell}$ decays starting from the corresponding $\Bbar \rightarrow D^{(*)} X \ell^- \bar{\nu}_{\ell}$ samples and selecting events with only one additional reconstructed charged track that has not been used for the reconstruction of the $B_{tag}$, the signal $D^{(*)}$, or the lepton. 
For the $\Bzb \rightarrow D^0\pi^+ \ell^- \bar{\nu}_{\ell}$ and the $\Bzb \rightarrow D^{*0}\pi^+ \ell^- \bar{\nu}_{\ell}$ decays, we additionally require the invariant mass difference $M(D\pi)-M(D)$ to be greater than 0.18 GeV/$c^2$ to veto $\Bzb \rightarrow D^{*+} \ell^- \bar{\nu}_{\ell}$ events. To reduce the combinatorial background in the $\Bzb \rightarrow D^{*0}\pi^+ \ell^- \bar{\nu}_{\ell}$ mode, we also require the total extra energy in the event, obtained by summing the energy of all the showers in the electromagnetic  calorimeter that have not been assigned to the $B_{tag}$ or the $D^{(*)}\ell^-$ candidates, to be less than 1 GeV.

The exclusive semileptonic $B$ decays are identified by the missing mass squared in the event, $m^2_{miss} = (p(\Upsilon(4S)) -p(B_{tag}) - p(D^{(*)}(\pi)) - p(\ell))^2$, defined in terms of the particle
four-momenta in the CM frame of the reconstructed final
states. For correctly reconstructed signal events, the only missing particle is the neutrino, and $m^2_{miss}$ peaks at zero. Other $B$ semileptonic decays, where one particle is not reconstructed (feed-down) or is erroneously added (feed-up) to the charm candidate, exhibit higher or lower values in $m^2_{miss}$. 
To obtain the $B$ semileptonic signal yields, we perform a one-dimensional extended binned maximum likelihood fit~\cite{Barlow} to the $m^2_{miss}$ distributions. The fitted data samples are assumed to contain four different types of events: $\Bbar \to D^{(*)}(\pi) \ell^- \bar{\nu}_{\ell}$ signal events, feed-down or feed-up from other $B$ semileptonic decays, combinatoric \BB\ and continuum background, and hadronic $B$ decays (mainly due to hadrons misidentified as leptons).
For the fit to the $m^2_{miss}$ distributions of the $\Bbar \to D^{(*)}\pi \ell^- \bar{\nu}_{\ell}$ channel, we also include a component corresponding to other misreconstructed $\Bbar \to D^{**}(D^{*}\pi) \ell^- \bar{\nu}_{\ell}$ decays. 
We use the MC predictions for the different $B$ semileptonic decay $m^2_{miss}$ distributions to obtain the Probability Density Functions (PDFs). The combinatoric \BB\ and continuum background shape is also estimated by the MC simulation, and we use the off-peak data to provide the continuum background normalization. The shape of the continuum background distribution predicted by the MC simulation is consistent with that obtained from the off-peak data. 

The $m^2_{miss}$ distributions are compared with the results of the fits in Fig.~1 for each of the $\Bbar \to D^{(*)}(\pi) \ell^- \bar{\nu}_{\ell}$ channels. The fitted signal yields and the signal efficiencies,  accounting for the $B_{tag}$ reconstruction, are listed in Table \ref{tab:results}.

\begin{figure}[!ht]
\includegraphics[scale=0.44]{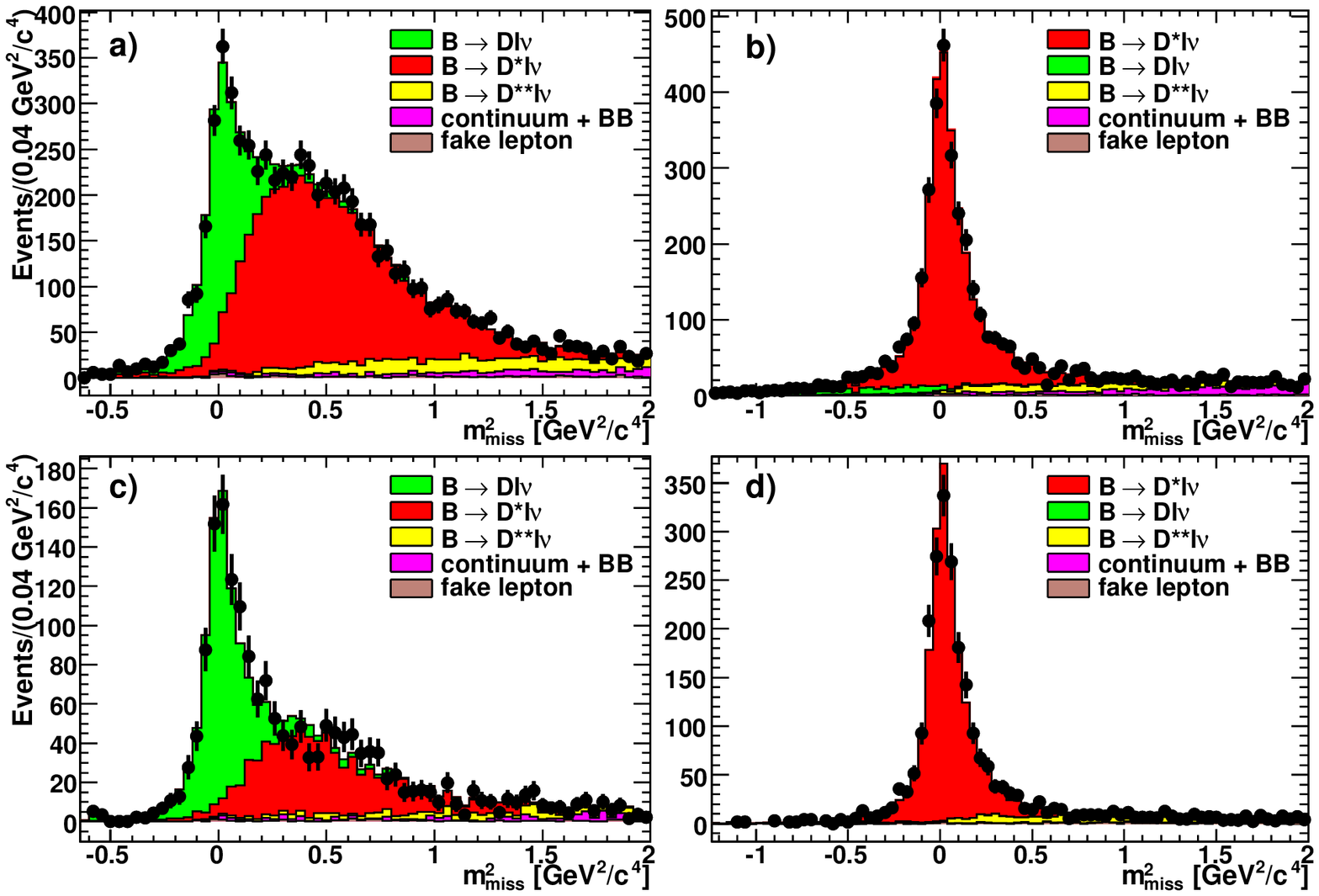}
\includegraphics[scale=0.44]{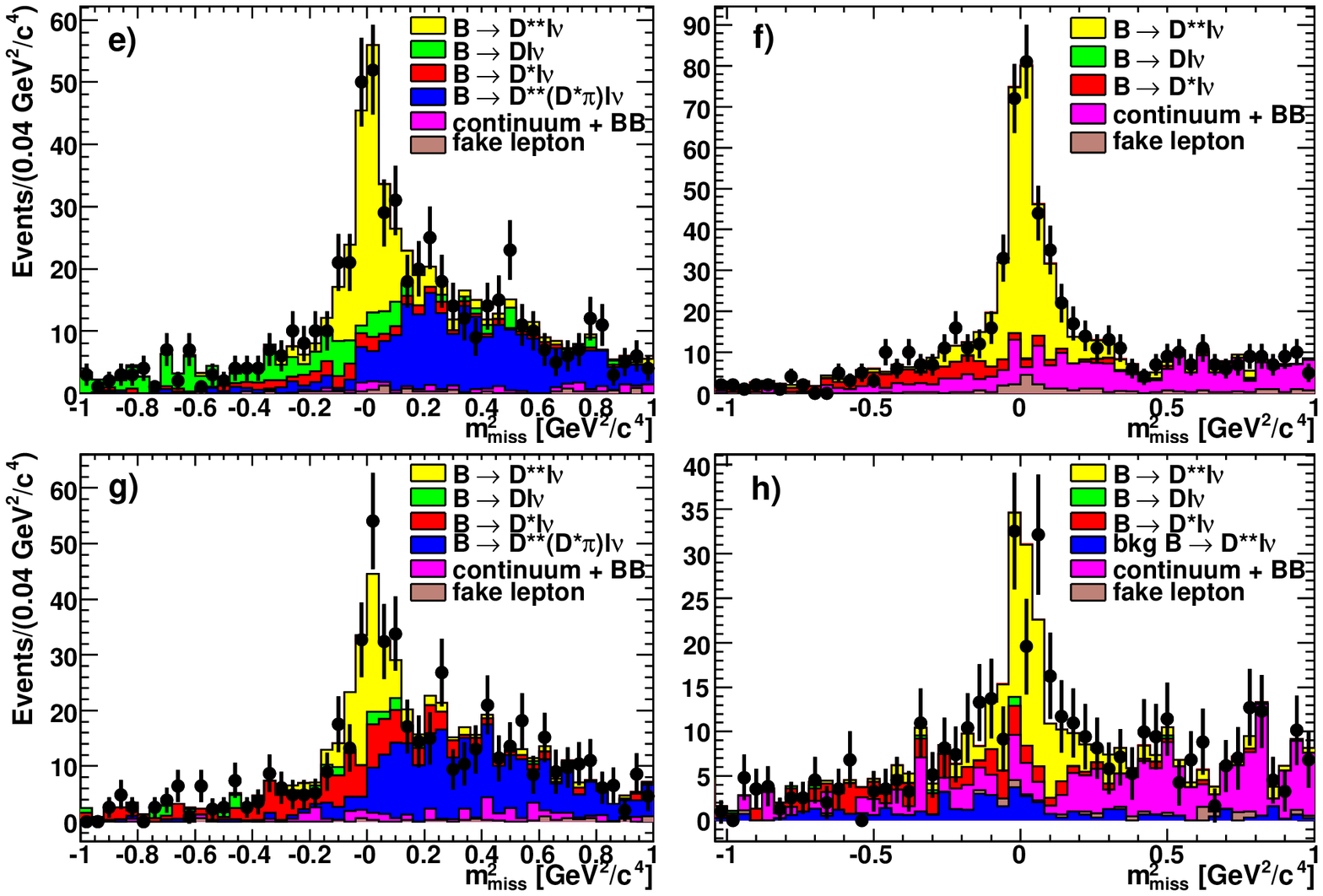}
\label{fig:Fit}
\caption{(Color online) Fit to the $m^2_{miss}$ distribution for a) $B^- \to D^0 \ell^- \bar{\nu}_{\ell}$, b) $B^- \to D^{*0} \ell^- \bar{\nu}_{\ell}$, c) $\Bzb \to D^+ \ell^- \bar{\nu}_{\ell}$, d) $\Bzb \to D^{*+} \ell^- \bar{\nu}_{\ell}$, e) $B^- \to D^+ \pi^- \ell^- \bar{\nu}_{\ell}$, f) $B^- \to D^{*+} \pi^- \ell^- \bar{\nu}_{\ell}$,  g) $\Bzb \to D^0 \pi^+ \ell^- \bar{\nu}_{\ell}$, and h) $\Bzb \to D^{*0} \pi^+ \ell^- \bar{\nu}_{\ell}$: the data (points with error bars) are compared to the results of the overall fit (sum of the solid histograms). The PDFs for the different fit components are stacked and shown in different colors.}
\end{figure}

\begin{figure}[!h]
\includegraphics[scale=0.2135]{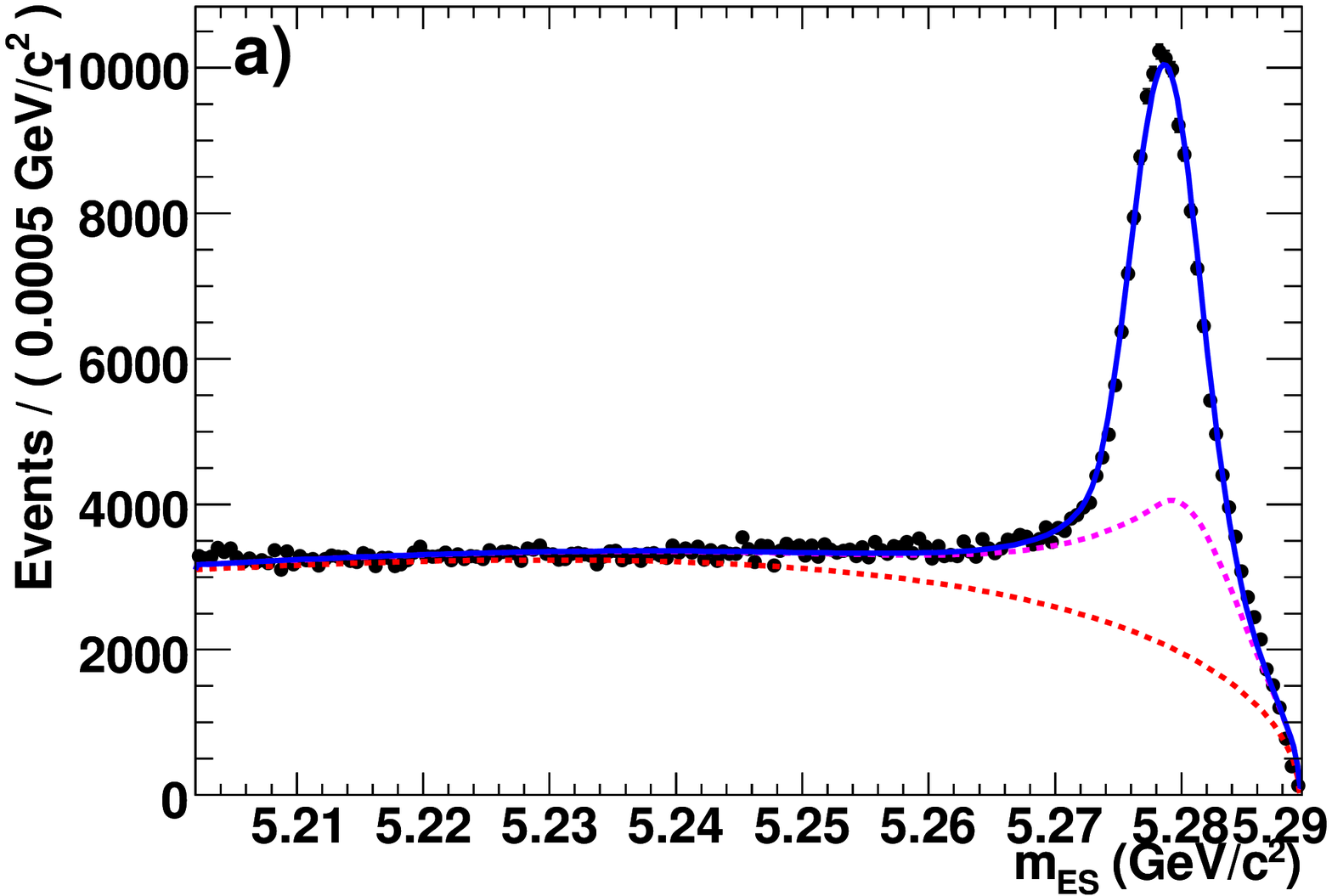}
\includegraphics[scale=0.2135]{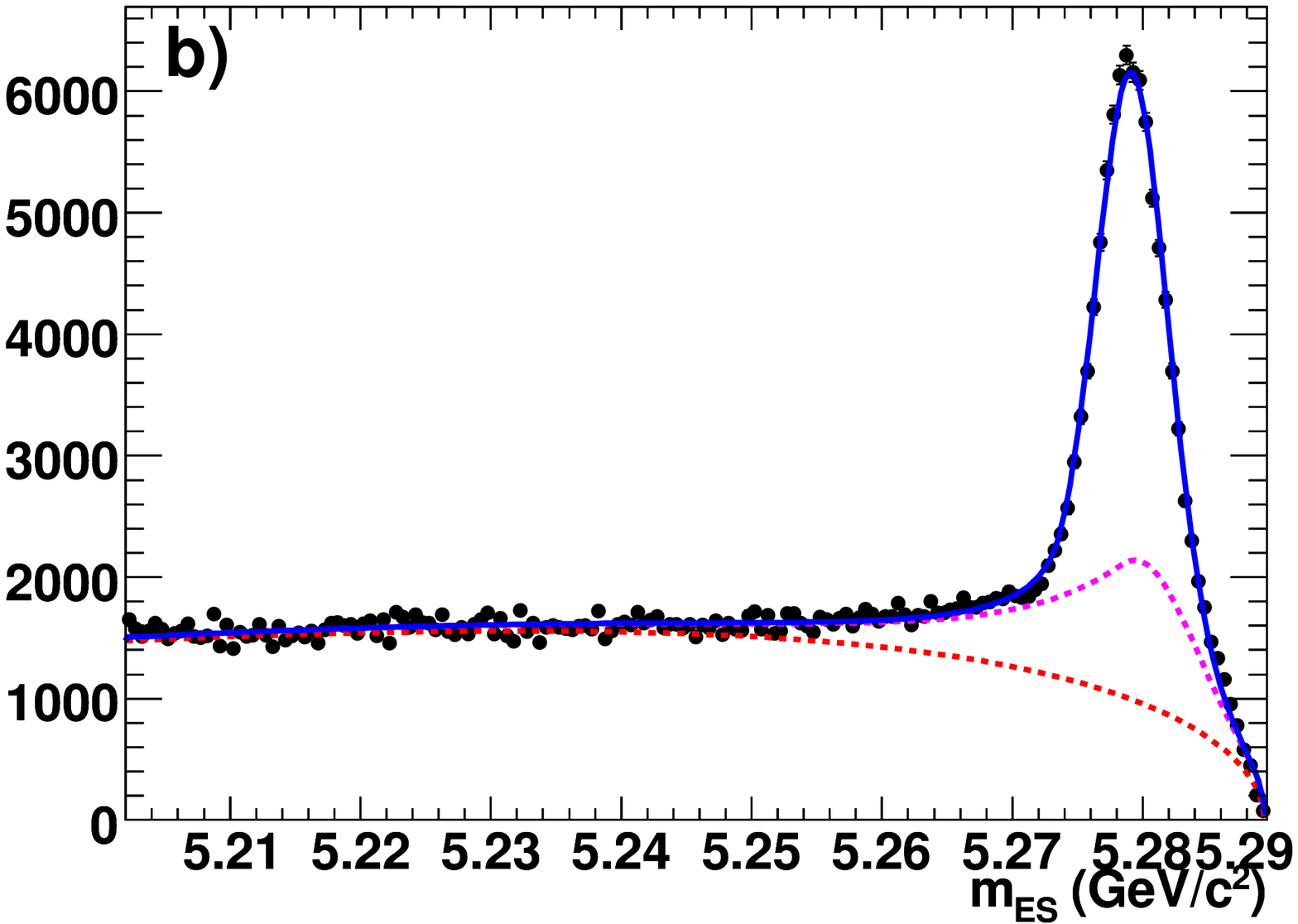}
\caption{(Color online) $m_{ES}$ distributions of the a) $B^- \to X \ell^- \bar{\nu}_{\ell}$, and b) $\Bzb \to X \ell^- \bar{\nu}_{\ell}$ samples. The data (points with error bars) are compared to the result of the fit (solid line). The dashed lines show the broad-peaking component and the sum of the combinatorial and continuum background.}
\label{fig:mesB}
\end{figure}

To reduce the systematic uncertainty, the exclusive ${\cal B} (\Bbar \rightarrow D^{(*)}(\pi) \ell^- \bar{\nu}_{\ell})$ branching fractions relative to the inclusive semileptonic branching fraction are measured. 
A sample of $\Bbar \to X \ell^- \bar{\nu}_{\ell}$ events is selected by identifying a charged lepton with CM momentum greater than 0.6 GeV/$c$ and the correct charge-flavor correlation with the $B_{tag}$ candidate. In the case of multiple $B_{tag}$ candidates in an event, we select the one reconstructed in the decay channel with the highest purity, defined as the fraction of signal events in the $m_{ES}$ signal region. 
Background components peaking in the $m_{ES}$ signal region include cascade $B$ meson decays ($i.e.$, the lepton does not come directly from the $B$) and hadronic decays, and are subtracted by using the corresponding MC  distributions. 
The total yield for the inclusive $\Bbar \to X \ell^- \bar{\nu}_{\ell}$ decays is obtained from a maximum likelihood fit to the $m_{ES}$ distribution of the $B_{tag}$ candidates using an ARGUS function~\cite{Argus} for the description of the combinatorial \BB\  and continuum background, and a Crystal Ball function~\cite{CrystallBall} for the signal. 
Additional Crystal Ball and ARGUS functions are used to model a broad-peaking component, included in the signal definition, due to real $\Bbar \to X \ell^- \bar{\nu}_{\ell}$ decays for which, in the $B_{tag}$ reconstruction, neutral particles have not been identified or have been interchanged with the semileptonic decays. 
Fig.~\ref{fig:mesB} shows the $m_{ES}$ distribution of the $B_{tag}$ candidates in the $B^- \to X \ell^- \bar{\nu}_{\ell}$ and $\Bzb \to X \ell^- \bar{\nu}_{\ell}$ sample. The fit yields 159896 $\pm$ 1361 events for the 
$B^- \to X \ell^- \bar{\nu}_{\ell}$ sample and 96771 $\pm$ 968 events for the $\Bzb \to X \ell^- \bar{\nu}_{\ell}$ sample. 

The relative branching fractions ${\cal B}(\Bbar \to D^{(*)}(\pi) \ell^- \bar{\nu}_{\ell})/{\cal B}(\Bbar \to X \ell^- \bar{\nu}_{\ell})$ are obtained by correcting the signal yields for the reconstruction efficiencies (estimated  from \BB\ MC events) and normalizing to the inclusive  $\Bbar \to X \ell^- \bar{\nu}_{\ell}$ signal yield, following the relation 
${\cal B}(\Bbar \rightarrow D^{(*)}(\pi) \ell^- \bar{\nu}_{\ell})/{\cal B}(\Bbar \to X \ell^- \bar{\nu}_{\ell}) = \frac{N_{sig}}{\epsilon_{sig}}\frac{\epsilon_{sl}}{N_{sl}}$. Here, $N_{sig}$ is the number of 
$\Bbar \to D^{(*)}(\pi) \ell^- \bar{\nu}_{\ell}$ signal events 
for the various modes, reported in  Table~\ref{tab:results} together
 with the corresponding reconstruction efficiencies 
$\epsilon_{sig}$, $N_{sl}$ is the $\Bbar \to X \ell^- 
\bar{\nu}_{\ell}$ signal yield, and $\epsilon_{sl}$ 
is the corresponding reconstruction efficiency 
including the $B_{tag}$ reconstruction, 
equal to 0.36\% and 0.23\% for the $B^- \to X \ell^- \bar{\nu}_{\ell}$ 
and $\Bzb \to X \ell^- \bar{\nu}_{\ell}$ decays, respectively. The absolute branching fractions ${\cal B} (\Bbar \rightarrow D^{(*)}(\pi) \ell^- \bar{\nu}_{\ell})$ are then determined using the semileptonic branching fraction ${\cal B}(\Bbar \to X \ell^- \bar{\nu}_{\ell})= ( 10.78 \pm 0.18)\%$ and the ratio of the $B^0$ and the $B^+$ lifetimes $\tau_{B^+}/\tau_{B^0} = 1.071 \pm 0.009$~\cite{pdg}.

\begin{table*}[!bht]
\caption{Signal yields and reconstruction efficiencies for the $\Bbar \to D^{(*)}(\pi) \ell^- \bar{\nu}_{\ell}$ decays. }
\begin{tabular}{l|c|c|l|c|c}
\hline
\hline
Decay Mode & $N_{sig}$ & $\epsilon_{sig} (\times 10^{-4})$ & Decay Mode & $N_{sig}$ & $\epsilon_{sig} (\times 10^{-4})$ \\
\hline
$B^- \rightarrow D^0 \ell^- \bar{\nu}_{\ell}$  & 1635 $\pm$ 61 & 1.71 $\pm$ 0.02 & $B^- \rightarrow D^+ \pi^- \ell^- \bar{\nu}_{\ell}$ & 174 $\pm$ 25 & 1.02 $\pm$ 0.03\\
\hline
$B^- \rightarrow D^{*0} \ell^- \bar{\nu}_{\ell}$  & 3050 $\pm$ 73 & 1.27 $\pm$ 0.01 & $B^- \rightarrow D^{*+} \pi^- \ell^- \bar{\nu}_{\ell}$ & 306 $\pm$ 27 & 1.26 $\pm$ 0.03\\
\hline
$\Bzb \rightarrow D^+ \ell^- \bar{\nu}_{\ell}$  & \hspace{0.07cm} 852 $\pm$ 40 & 0.94 $\pm$ 0.02 & $\Bzb \rightarrow D^0 \pi^+ \ell^- \bar{\nu}_{\ell}$ & 107 $\pm$ 20 & 0.60 $\pm$ 0.03\\
\hline
$\Bzb \rightarrow D^{*+}\ell^- \bar{\nu}_{\ell}$  & 2045 $\pm$ 55  & 0.91 $\pm$ 0.01 & $ \Bzb \rightarrow D^{*0} \pi^+ \ell^- \bar{\nu}_{\ell}$ & 130 $\pm$ 20 & 0.66 $\pm$ 0.02\\
\hline
\hline
\end{tabular}
\label{tab:results}
\end{table*}

Numerous sources of systematic uncertainties have been investigated.
The uncertainties due to the detector simulation are established by varying, within bounds given by data control samples, the tracking efficiency of all charged tracks (resulting in 1.2-2.7\% relative systematic uncertainty among the different decay modes), the calorimeter efficiency (0.5-1.8\%), the lepton identification efficiency (0.4-3\%),  and the reconstruction efficiency for low momentum charged (1.2\%) and neutral pions (1.3\%).  
We evaluate the systematic uncertainties associated with the MC simulation of various signal and background processes: photon conversion and $\pi^0$ Dalitz decay (0.04-0.4\%), $B$ cascade decay contamination (0.6-1\%), and  flavor cross-feed (0.2-0.3\%). 
We vary the $\Bbar \to D \ell^- \bar{\nu}_{\ell}$ and  $\Bbar \to D^{*} \ell^- \bar{\nu}_{\ell}$ form factors within their measured uncertainties~\cite{HQETBaBar} (0.4-0.8\%) and we include the uncertainty on the branching fractions of the reconstructed $D$ and $D^{*}$ modes (2.3-4.4\%), and on the absolute branching fraction ${\cal B} (\Bbar \to X \ell^- \bar{\nu}_{\ell})$ used for the normalization (1.9\%).
We also include a systematic uncertainty due to differences in the efficiency of the $B_{tag}$ selection in the exclusive selection of $\Bbar \to D^{(*)}(\pi) \ell^- \bar{\nu}_{\ell}$ decays and the inclusive $\Bbar \to X \ell^- \bar{\nu}_{\ell}$ reconstruction (0.9-5.6\%), and the extraction of the $\Bbar \to D^{(*)}(\pi) \ell^- \bar{\nu}_{\ell}$ (0.4-1.8\%) and $\Bbar \to X \ell^- \bar{\nu}_{\ell}$ (0.5-0.9\%) signal yields.
The complete set of systematic uncertainties is given in Ref.~\cite{epaps}.

We measure the following branching fractions:

\begin{eqnarray}
{\cal B} (B^- \rightarrow D^0 \ell^- \bar{\nu}_{\ell})  &=& (2.33 \pm 0.09_{stat.} \pm 0.09_{syst.}) \%  \nonumber \\
{\cal B} (B^- \rightarrow D^{*0} \ell^- \bar{\nu}_{\ell}) &=& (5.83 \pm 0.15_{stat.} \pm 0.30_{syst.}) \% \nonumber \\
{\cal B} (\Bzb \rightarrow D^+ \ell^- \bar{\nu}_{\ell})&=& (2.21 \pm 0.11_{stat.} \pm 0.12_{syst.}) \% \nonumber \\
{\cal B} (\Bzb \rightarrow D^{*+}\ell^- \bar{\nu}_{\ell}) &=& (5.49 \pm 0.16_{stat.} \pm 0.25_{syst.}) \% \nonumber \\
{\cal B} (B^- \rightarrow D^+ \pi^- \ell^- \bar{\nu}_{\ell}) &=& (0.42 \pm 0.06_{stat.} \pm 0.03_{syst.} ) \% \nonumber \\
{\cal B} (B^- \rightarrow D^{*+} \pi^- \ell^- \bar{\nu}_{\ell}) &=& (0.59 \pm 0.05_{stat.} \pm 0.04_{syst.}) \% \nonumber \\
{\cal B} (\Bzb \rightarrow D^0 \pi^+ \ell^- \bar{\nu}_{\ell})&=& (0.43 \pm 0.08_{stat.} \pm 0.03_{syst.}) \% \nonumber \\
{\cal B} (\Bzb \rightarrow D^{*0} \pi^+ \ell^- \bar{\nu}_{\ell}) &=& (0.48 \pm 0.08_{stat.} \pm 0.04_{syst.}) \%. \nonumber \\ \nonumber
\end{eqnarray}

The accuracy of the branching fraction measurements for the $\Bbar \rightarrow D^{(*)} \ell^- \bar{\nu}_{\ell}$ decays is comparable to that of the current world average~\cite{pdg}. We compute the total branching fractions of the $\Bbar \rightarrow D^{(*)} \pi \ell^- \bar{\nu}_{\ell}$ decays assuming isospin symmetry, ${\cal B} (\Bbar \rightarrow D^{(*)}\pi^0 \ell^- \bar{\nu}_{\ell})=\frac{1}{2} {\cal B} (\Bbar \rightarrow D^{(*)}\pi^{\pm} \ell^- \bar{\nu}_{\ell})$, to estimate the branching fractions of $D^{(*)} \pi^0$ final states, obtaining:

\begin{eqnarray}
{\cal B} (B^- \rightarrow D^{(*)}\pi \ell^- \bar{\nu}_{\ell}) &=& (1.52 \pm  0.12_{stat.} \pm 0.10_{syst.}) \% \nonumber \\ 
{\cal B} (\Bzb \rightarrow D^{(*)}\pi \ell^- \bar{\nu}_{\ell}) &=& (1.37 \pm  0.17_{stat.} \pm 0.10_{syst.}) \%,  \nonumber \\ \nonumber 
\end{eqnarray}

\noindent where we assume the systematic uncertainties on the $\Bbar \rightarrow D \pi \ell^- \bar{\nu}_{\ell}$ and $\Bbar \rightarrow D^{*} \pi \ell^- \bar{\nu}_{\ell}$ modes to be completely correlated. These results are consistent  with, but have smaller uncertainties than, recent results from the Belle collaboration~\cite{belle}.

By comparing the sum of the measured branching fractions for $\Bbar \to D^{(*)}(\pi) \ell^- \bar{\nu}_{\ell}$ with the inclusive $\Bbar \to X_c \ell^- \bar{\nu}_{\ell}$ branching fraction~\cite{pdg}, a $(11 \pm 4)\%$ discrepancy is observed, which is most likely due to $\Bbar \to D^{(*)}n \pi \ell^- \bar{\nu}_{\ell}$ decays with $n>1$. 

We are grateful for the excellent luminosity and machine conditions
provided by our \pep2\ colleagues, 
and for the substantial dedicated effort from
the computing organizations that support \babar.
The collaborating institutions wish to thank 
SLAC for its support and kind hospitality. 
This work is supported by
DOE
and NSF (USA),
NSERC (Canada),
CEA and
CNRS-IN2P3
(France),
BMBF and DFG
(Germany),
INFN (Italy),
FOM (The Netherlands),
NFR (Norway),
MIST (Russia),
MEC (Spain), and
STFC (United Kingdom). 
Individuals have received support from the
Marie Curie EIF (European Union) and
the A.~P.~Sloan Foundation.

\clearpage

\begin{table*}
\flushleft
\textbf{\large Electronic Physics Auxiliary Publication Service (EPAPS)}\\
\smallskip
\normalsize
This is an EPAPS attachment to B.~Aubert \textit{et al.} (\babar\ Collaboration),
\babar-PUB-\BABARPubYear/\BABARPubNumber, SLAC-PUB-\SLACPubNumber, 
submitted to Phys.\ Rev.\ Lett. For more information on EPAPS, see
http://www.aip.org/pubservs/epaps.html.
\end{table*}

\begin{table*}
\centering
\caption{Systematic uncertainties (relative errors in \%) in the measurement of ${\cal B}(\overline{B} \rightarrow D^{(*)} \ell^- \bar{\nu}_{\ell})/{\cal B}(\overline{B} \rightarrow X \ell^- \bar{\nu}_{\ell})$.}
\begin{tabular}{|c|c|c|c|c|}
\hline
\hline
 &\multicolumn{4}{c|}{{\scriptsize Systematic uncertainty on ${\cal B}(\overline{B} \rightarrow D^{(*)}\ell^- \bar{\nu}_{\ell}) /{\cal B}(\overline{B} \rightarrow X \ell^- \bar{\nu}_{\ell})$} } \\
\hline
& $B^- \rightarrow D^0\ell^-\bar{\nu}_{\ell}$ & $B^- \rightarrow D^{*0}\ell^-\bar{\nu}_{\ell}$ & $\overline{B^0} \rightarrow D^{+}\ell^-\bar{\nu}_{\ell}$ & $\overline{B^0} \rightarrow D^{*+} \ell^-\bar{\nu}_{\ell}$ \\
\hline
Tracking efficiency & 1.4 & 1.2 & 1.4 & 1.5 \\
Neutral reconstruction & 0.7  & 1.9 & 0.5 & 1.1 \\
lepton ID & 0.5  & 0.4  & 0.5 & 0.6 \\
Soft particle efficiency & -  & 1.3 & -  & 1.2\\
\hline
Monte Carlo corrections & \multicolumn{4}{c|}{} \\
\hline
Conversion and Dalitz decay background & 0.04 & 0.07  & 0.06 & 0.05\\
Cascade $\overline{B} \to X \to \ell^-$ decay background  & 0.6 & 0.6 & 1.0 & 1.0\\
$\overline{B^0}-B^-$ cross-feed & 0.2 & 0.3 & 0.2 & 0.3\\
Form factors & 0.4 & 0.8 & 0.4 & 0.8 \\
$D$ branching fractions & 2.3  & 2.1 & 4.1  & 2.6\\
$D^*$ branching fractions & -  & 2.8 & -  & 0.8\\
\hline
$\overline{B} \to X \ell^- \bar{\nu}_{\ell}$ branching fraction & 1.9  & 1.9 & 1.9  & 1.9\\
\hline
$B_{tag}$ selection & 0.9 & 1.7 & 1.8 & 1.3\\
\hline
Fit technique & \multicolumn{4}{c|}{}\\
\hline
$\overline{B} \to X \ell^- \bar{\nu}_{\ell}$ yield & 0.5 & 0.5 & 0.9 & 0.9\\
$\overline{B} \rightarrow D^{(*)} \ell^- \bar{\nu}_{\ell}$ yield & 0.6 & 0.4 & 1.2 & 0.4\\
\hline
Total systematic error & 3.7 & 5.2 & 5.4 & 4.5\\
\hline
\hline
\end{tabular}
\label{tab:Syst1}
\end{table*}

\begin{table*}
\centering
\caption{Systematic uncertainties (relative errors in \%) in the measurement of ${\cal B}(\overline{B} \rightarrow D^{(*)}\pi \ell^- \bar{\nu}_{\ell})/{\cal B}(\overline{B} \rightarrow X \ell^- \bar{\nu}_{\ell})$.}
\begin{tabular}{|c|c|c|c|c|}
\hline
\hline
 &\multicolumn{4}{c|}{{\scriptsize Systematic uncertainty on ${\cal B}(\overline{B} \rightarrow D^{(*)} \pi \ell^- \bar{\nu}_{\ell}) /{\cal B}(\overline{B} \rightarrow X \ell^- \bar{\nu}_{\ell})$} } \\
\hline
&  $B^- \rightarrow D^{+}\pi^- \ell^- \bar{\nu}_{\ell}$ & $B^- \rightarrow D^{*+} \pi^- \ell^- \bar{\nu}_{\ell}$ & $\overline{B^0} \rightarrow D^0 \pi^+ \ell^-\bar{\nu}_{\ell}$ & $\overline{B^0} \rightarrow D^{*0} \pi^+ \ell^- \bar{\nu}_{\ell}$  \\
\hline
Tracking efficiency  & 1.8 & 2.7 & 1.5 & 1.7\\
Neutral reconstruction  & 1.7 & 1.8 & 1.1 & 1.8\\
lepton ID  & 2.3 & 3.0 & 2.6  & 1.8\\
Soft particle efficiency  &  - & 1.2 & -  & 1.3\\
\hline
Monte Carlo corrections & \multicolumn{4}{c|}{} \\
\hline
Conversion and Dalitz decay background  & 0.15 & 0.4 & 0.05 & 0.2\\
Cascade $\overline{B} \to X \to \ell^-$ decay background   & 0.6 & 0.6 & 1.0 & 1.0\\
$\overline{B^0}-B^-$ cross-feed  & 0.2 & 0.3 & 0.2  & 0.3\\
Form factors  & 0.4 & 0.8 & 0.4 & 0.8\\
$D$ branching fractions  & 4.2 & 2.8 & 2.5 & 2.9\\
$D^*$ branching fractions  & - & 0.9 & - & 3.3\\
\hline
$\overline{B} \to X \ell^- \bar{\nu}_{\ell}$ branching fraction & 1.9  & 1.9 & 1.9  & 1.9\\
\hline
$B_{tag}$ selection  & 5.0 & 4.3 & 4.0 & 5.6\\
\hline
Fit technique & \multicolumn{4}{c|}{}\\
\hline
$\overline{B} \to X \ell^- \bar{\nu}_{\ell}$ yield   & 0.5 & 0.5 & 0.9 & 0.9\\
$\overline{B} \rightarrow D^{(*)} \pi \ell^- \bar{\nu}_{\ell}$ yield  & 1.2 & 0.9 & 1.8 & 1.5 \\
\hline
Total systematic error  & 7.7 &  7.3 & 6.4 & 8.4 \\
\hline
\hline
\end{tabular}
\label{tab:Syst2}
\end{table*}


\begin{thebibliography}{99}
\bibitem{ell}
Here $X_c$ refers to any charm hadronic state, $X_u$ to any charmless hadronic state, $X=X_c+X_u$ and $\ell = e, \mu$. The charge conjugate state is always implied unless stated otherwise. 
\bibitem{CKM}
M.~Kobayashi and T.~Maskawa, Prog.~Theor.~Phys. {\bf 49},~652 (1973).
\bibitem{aleph}
D.~Buskulic {\it et al.} (ALEPH Collab.), Z.~Phys.~C {\bf 73},~601 (1997).
\bibitem{delphi}
P.~Abreu {\it et al.} (DELPHI Collab.), Phys.~Lett.~B {\bf 475},~407 (2000).
\bibitem{babar-1}
B.~Aubert {\it et al.} (\babar\ Collab.), Phys.~Rev.~D{\bf 71},~051502 (2005).
\bibitem{babar-2}
B.~Aubert {\it et al.} (\babar\ Collab.), Phys.~Rev.~D{\bf 76},~051101 (2007).
\bibitem{belle}
D.~Liventsev {\it et al.} (Belle Collab.), Phys.~Rev.~D{\bf 72},~051109 (2005).
\bibitem{pdg}
W.~-M.~Yao {\it et al.}  (Particle Data Group), J.~Phys.~G {\bf 33},~1 (2006).
\bibitem{detector}
B.~Aubert {\it et al.} (\babar\ Collab.), Nucl.~Instrum.~Methods A{\bf 479},~1 (2002).
\bibitem{Geant}
S.~Agostinelli {\it et al.}, Nucl.~Instrum.~Methods A{\bf 506},~250 (2003).
\bibitem{HQETBaBar}
I.~Caprini, L.~Lellouch and M.~Neubert, Nucl.~Phys.~B{\bf 505},~65 (1997); B.~Aubert {\it et al.} (\babar\ Collaboration), arXiv:0705.4008 [hep-ex], submitted to PRD.
\bibitem{ISGW}
D.~Scora and N.~Isgur, Phys.~Rev.~D{\bf 52},~2783 
(1995). See also N.~Isgur {\it et al.}, Phys.~Rev.~D{\bf 39},~799 (1989).
\bibitem{Goity}
L.~Goity and W.~Roberts, Phys.~Rev.~D{\bf 51},~3459 (1995).
\bibitem{BBmixing}
B.~Aubert {\it et al.} (\babar\ Collaboration), Phys.~Rev.~D{\bf 69},~111104 (2004).
\bibitem{Barlow}
R.~J.~Barlow and C.~Beeston, Comput.~Phys.~Commun.~{\bf 77}, 219 (1993).
\bibitem{Argus}
H.~Albrecht {\it et al.} (ARGUS Collaboration), Z.~Phys.~C{\bf 48},~543 (1990).
\bibitem{CrystallBall}
M.~J.~Oreglia, SLAC-236~(1980);
J.~E.~Gaiser, SLAC-255~(1982);
T.~Skwarnicki, DESY F31-86-02~(1986).
\bibitem{epaps}
See EPAPS Document No. $xxxxx$ for the complete set of systematic uncertainties. For more information on EPAPS, see http://www.aip.org/pubservs/epaps.html.
\end{thebibliography}
\end{document}